\documentclass[preprint,onecolumn,showpacs,preprintnumbers,amsmath,amssymb]{revtex4}
\usepackage{epsf}
\usepackage{dcolumn}
\usepackage{bm}
\everymath{\displaystyle}

\begin{document}

\title{Spin precession of a particle with an electric dipole moment: contributions from classical electrodynamics
and from the Thomas effect}

\author{Alexander J. Silenko}

\affiliation{Research Institute for Nuclear Problems, Belarusian State University, Minsk 220030, Belarus\\
Bogoliubov Laboratory of Theoretical Physics, Joint Institute for Nuclear Research,
Dubna 141980, Russia}

\date{\today}

\begin{abstract}
The new derivation of the equation of the spin precession is given for a particle possessing electric and magnetic
dipole moments. Contributions from classical electrodynamics and from the Thomas effect are explicitly separated. A fully covariant approach is used. The final equation is expressed in a very simple form in terms of the fields in the instantaneously accompanying frame.
The Lorentz transformations of the electric and magnetic dipole moments and of the spin are derived from basic equations of classical electrodynamics.
For this purpose, the Maxwell equations in matter are used and the result is confirmed by other methods. An antisymmetric four-tensor is correctly constructed from the electric and magnetic dipole moments.
\end{abstract}

\maketitle

\section{Introduction}

A spin motion of a particle with an anomalous magnetic
moment (AMM) and an electric dipole moment (EDM) in electric and magnetic fields is an important problem of classical spin physics.
A search for the EDM \cite{EDMrevs} is a part of exploration of new physics beyond the Standard Model. For charged particles in storage rings, this search is based on the relativistic equation of spin motion. The corresponding equation for a particle without the EDM has been derived by Thomas \cite{Thomas} (also by Frenkel \cite{F}) and, in a more general form, by Bargmann, Michel and
Telegdi \cite{BMT}. This is so-called Thomas-Bargmann-Michel-Telegdi (T-BMT) equation. There are two main methods of derivation of this equation. The Thomas method \cite{Thomas} (clearly explained in Ref. \cite{MShY}) is based on separated calculations of the spin precession in the instantaneously accompanying frame and of the contribution from the Thomas effect. Addition of this contribution to the angular velocity of the spin precession obtained with a Lorentz transformation from the instantaneously accompanying frame leads to the needed equation. The Bargmann-Michel-Telegdi method \cite{Thomas} (transparently clarified in Ref. \cite{Landau}) consists in the use of covariant equations of motions for the four-vectors of spin and velocity, $a^\mu$ and $u^\mu$, and the orthogonality condition $a_\mu u^\mu=0$. The transition to the rest frame spin, $\bm\zeta$, allows one to derive the T-BMT equation. This method does not explicitly use the equation for the angular velocity of the Thomas precession.

An extension of the T-BMT equation due to the EDM has already been discussed in the original paper of Bargmann, Michel and Telegdi \cite{BMT}. Then, the equation of spin motion of the particle with the AMM, $\mu'$, and the EDM, $d$, has been obtained in Refs. \cite{Nelson,Khriplovich} by the dual
transformation
$\mu'\rightarrow d,~{\bm B}\rightarrow{\bm E},~{\bm E}\rightarrow-{\bm B}$.
The rigorous derivation of this equation has been presented in Ref. \cite{FukuyamaSilenko}. The resulting equation of spin motion coincides with that presented in Refs. \cite{Nelson,Khriplovich}. However, the derivation fulfilled in Ref. \cite{FukuyamaSilenko} has not used the supplementary assumption of dual symmetry.

We demonstrate in the present work that including the EDM into a consideration opens new possibilities to relate the particle spin motion with basic equations of classical electrodynamics, namely,
the Maxwell equations in matter and the Lorentz transformations of the four-current and other four-vectors.
We also extract a contribution from the Thomas effect to the resulting spin motion with the use of a fully covariant approach.

\section{Electromagnetic interactions of a moving particle}\label{Lorentz}

We consider an extended charged particle in electric and magnetic fields. In fact, the fields may be nonuniform and nonstationary if we 
neglect terms proportional to their derivatives. In the framework of classical electrodynamics, we can divide the particle into point-like charges $e$ and currents $\bm J$.
Let $\bm R$ be the radius-vector of the center of mass of the particle:
\begin{equation}  \bm R=\frac{\sum{\mathfrak{E}\bm r}}{\sum{\mathfrak{E}}}, \label{CoM} \end{equation} where $\mathfrak{E}$ and $\bm r$ are the total energy and the radius-vector
of a constituent part of the particle.

One defines an interaction of the electric and magnetic dipole moments, $\bm d$ and $\bm\mu$, with the external fields by the Hamiltonian
\begin{equation}   \begin{array}{c}
H=-\bm d\cdot\bm E-\bm\mu\cdot\bm B,
\end{array} \label{multexp} \end{equation} where
\begin{equation}   \begin{array}{c} \bm d=\sum{e\bm r}=\int{\rho(\bm r)\bm rdV},
\end{array} \label{dentrmq} \end{equation}
\begin{equation}   \begin{array}{c}
\bm\mu=\frac{1}{2c}\sum{[\bm r\times\bm J]}=\frac{1}{2c}\int{[\bm
r\times\bm j]dV}.
\end{array} \label{mdipold} \end{equation}
Here $\rho(\bm r)$ and $\bm j(\bm r)$ are the charge and current densities. The sums are replaced with the integrals.
This conventional definition becomes inexact for the moving particle. The rigorous definition should take into account a motion of the center of mass $\bm R(t)$. When its velocity is $\bm V$, the EDM takes the form
\begin{equation}  \bm d=\sum{e\bm{\mathfrak{r}}}, ~~~ \bm{\mathfrak{r}}=\bm r-\bm R(0)-\bm Vt. \label{dCoM} \end{equation} 
Similar correction should be made for the magnetic moment:
\begin{equation}   \begin{array}{c}
\bm\mu=\frac{1}{2c}\sum{[\bm{\mathfrak{r}}\times\bm J]}=\frac{1}{2c}\int{[\bm{\mathfrak{r}}\times\bm j]dV}.
\end{array} \label{mdipd} \end{equation}

The center-of-mass velocity is given by
\begin{equation}  \bm V=\frac{c^2\sum{\bm\pi}}{\sum{\mathfrak{E}}}=\frac{c\sum{\bm\pi}}{\sum{\sqrt{\mathfrak{M}^2c^2+\bm\pi^2}}}, \label{CoMvelo} \end{equation}
where $\mathfrak{M}$ is the mass and $\bm\pi$ is the kinetic momentum of a constituent part of the particle.

Equation (\ref{dCoM}) defines the EDM relative to the center of mass. It is important to consider any another definition. The definition of the EDM relative to the center of charge $\bm R_c(t)$ results in
\begin{equation}  \bm d_c=\sum{e[\bm r-\bm R_c(0)-\bm V_ct]} \label{dCoz} \end{equation}
instead of Eq. (\ref{dCoM}). However, the difference between Eqs.  (\ref{dCoM}) and (\ref{dCoz}) is not very important.
Detailed classical description of an extended spinning particle has been presented in Ref. \cite{Rivas}. The spin and the averaged magnetic moment of such a particle are collinear. The electric and magnetic dipole moments are orthogonal in the particle rest frame. Therefore, the EDM oscillates and its average value in the particle rest frame is equal to zero. More exactly, it can be nonzero only when the \emph{CP} invariance is violated. These properties are valid for any definition of $\bm d$ and $\bm\mu$.
When we neglect the \emph{CP} violation, $<\bm d>=0$ and $<\bm d_c>=0$ \emph{in the particle rest frame}. As a result, their averaged difference is also equal to zero and $<\bm R_c>=\bm R$. For example, the center of a proton in a deuteron is equal to $\bm R_c$ and the proton rotates and oscillates relative to the center of mass. 
The \emph{CP} violation changes nothing because its influence on an atomic and nuclear structure can be neglected.

In the present paper, we use the standard definition of the EDM relative to the center of mass.

Electric and magnetic dipole moments of the particle depend on a reference frame.
In this section, we find a connection between the dipole moments in the lab frame and in the instantaneously
accompanying one. The connection between the latter frame and the rest frame (which is noninertial) has been found by Thomas \cite{Thomas}.

To express the electromagnetic interactions of the moving particle in terms of the intrinsic dipole moments, we can use the covariant form of the well-known expression for the relativistic
transformation of lengths:
\begin{equation}
\mathfrak{r}_{i}=\mathfrak{r}_{i}^{(0)}-\frac{\gamma}{\gamma+1}\beta_{i}\beta_{k}
\mathfrak{r}_{k}^{(0)}, ~~~\beta_i=\frac{V_{i}}{c}=\frac1c\cdot\frac{dX_{i}}{dt},
\label{eqpapYF} \end{equation} where $X_{i}$ are components of $\bm R$ and $\mathfrak{r}_{i}^{(0)}$ relate to the instantaneously accompanying frame.

A motion of the magnetic moment leads to the appearance of the EDM and other way round. Since the charge and current densities form a four-vector, the charge density is influenced by the motion of currents constituting a magnetic dipole:
\begin{equation}
\rho=\frac{\gamma}{c}\bm \beta\cdot\bm j^{(0)}. \label{meqFurl} \end{equation}
The current EDM \cite{MLCLE} appearing due to a motion of the magnetic dipole is given by $\bm d=\bm \beta\times\bm \mu^{(0)}$, where $\bm \mu^{(0)}$ is the magnetic moment in the instantaneously accompanying frame.

To derive general equations for the dipole moments and for the Hamiltonian of the moving particle, it
is convenient to use the Maxwell equations in matter (specifically, the Lorentz transformations of the Maxwell fields).
It is possible to determine physical quantities which are based on
the electric and magnetic dipole moments and form an antisymmetric
four-tensor. These are the electric and magnetic dipole moment densities
$$\bm P=\frac{d\bm d}{dV},~~~ \bm M=\frac{d\bm\mu}{dV}.$$
They enter into the Maxwell
equations in matter:
\begin{equation}\begin{array} {c}
\nabla\times\bm E=-\frac 1c\cdot\frac{\partial\bm B}{\partial t},~~~\nabla\times\bm H=\frac{4\pi}{c}\bm j^{(ext)}+\frac 1c\cdot\frac{\partial\bm D}{\partial t},\\
\nabla\cdot{\bm D}=4\pi\rho^{(ext)},~~~\nabla\cdot{\bm B}=0,~~~\bm D=\bm E+4\pi\bm P,~~~\bm B=\bm H+4\pi\bm M,
\end{array}
\label{Maxweqim}\end{equation} where $\rho^{(ext)}$ and $\bm
j^{(ext)}$ are the densities of external charges and currents. As
a result, $\bm P$ and $\bm M$ transform like the electric and
magnetic fields, $\bm E$ and $\bm B$. Since $F^{\mu\nu}=(-\bm E,\bm B)$ and $\mathfrak{G}^{\mu\nu}=(-\bm D,\bm H)$ are antisymmetric
four-tensors, the electric and magnetic dipole moment densities also form the antisymmetric
four-tensor ${\cal P}^{\mu\nu}=(\bm P,\bm M)$ (\cite{Jackson}, p. 557).

The Lorentz contraction defined by Eq. (\ref{eqpapYF}) results in
\begin{equation}
dV=\frac{dV^{(0)}}{\gamma}.
\label{eqvolum}\end{equation}
Therefore, the electric and magnetic dipole moments do not form a four-tensor and the related four-tensor has the form
\begin{equation}\begin{array} {c}
{\cal D}^{\mu\nu}=(\gamma\bm d,\gamma\bm \mu).
\end{array}
\label{eqKLnmn}\end{equation}

It is not evident that Eq. (\ref{eqvolum}) ensures the Lorentz invariance of charge. The elementary charge is given by $de=\rho dV$, where the charge density $\rho$ is a
zeroth component of a four-vector. When there are not any currents in the instantaneously accompanying frame ($\bm j^{(0)}=0$), the lab frame charge density is equal to $\rho=\gamma\rho^{(0)}$. As a result, $de=de^{(0)}$. Certainly, the current $\bm j=c\bm\beta\rho=c\bm\beta\gamma\rho^{(0)}$ also appears in the lab frame. When there are only currents in the instantaneously accompanying frame ($\bm\rho^{(0)}=0,~\bm j^{(0)}\neq0$), the lab frame charge density is given by Eq. (\ref{meqFurl}). Nevertheless, this does not violate the Lorentz invariance of charge. Any real current is closed. For a closed current, the sum of all positive charges appearing in the lab frame is equal (with an opposite sign) to the sum of all negative charges.

The use of the well-known transformation law for four-tensors leads to the following equations ($\gamma^{(0)}=1$):
\begin{equation}
\bm d=\bm d^{(0)}-\frac{\gamma}{\gamma+1}\bm\beta(\bm\beta\cdot\bm d^{(0)})
+\bm \beta\times\bm \mu^{(0)},
\label{deqfinal} \end{equation}
%
\begin{equation}
\bm\mu=\bm\mu^{(0)}-\frac{\gamma}{\gamma+1}\bm\beta(\bm\beta\cdot\bm\mu^{(0)})
-\bm \beta\times\bm d^{(0)}.
\label{meqfinal} \end{equation}
It is important that the
Lorentz transformation of the dipole moments can be connected with
the Maxwell equations. Equations (\ref{deqfinal}) and
(\ref{meqfinal}) can be obtained when $\bm P$ and $\bm M$ are
small as compared with $\bm E$ and $\bm B$, respectively.

Equations (\ref{deqfinal}) and (\ref{meqfinal}) can also be
derived by other methods.
Let us consider a particle possessing a
magnetic moment in the instantaneously accompanying frame. To obtain the
two first terms in the right hand side of Eq. (\ref{meqfinal}), we
can use the fact that the quantity $\bm{\mathfrak{r}}\times\bm j$ is the vector product of spatial
parts of two four-vectors. As a result, its Lorentz
transformation is similar to that of the angular momentum. This
transformation written in a covariant form
is given by
$$\bm{\mathfrak{r}}\times\bm j=\gamma\left\{[\bm{\mathfrak{r}}\times\bm j]^{(0)}-\frac{\gamma}{\gamma+1}\bm\beta(\bm\beta\cdot[\bm{\mathfrak{r}}\times\bm j]^{(0)})\right\}.$$
Taking into account Eqs. (\ref{mdipd}) and (\ref{eqvolum}) allows us to obtain the right expression
\begin{equation}
\bm\mu=\bm\mu^{(0)}-\frac{\gamma}{\gamma+1}\bm\beta(\bm\beta\cdot\bm\mu^{(0)}).
\label{meqFNAL} \end{equation}

To derive the last term in Eq. (\ref{deqfinal}), we can take into
account the Lorentz transformation of the four-current
(\ref{meqFurl}). Equations (\ref{dCoM}), (\ref{eqpapYF}), and
(\ref{meqFurl}) result in
\begin{equation}
\bm d=\frac{1}{c}\int{(\bm\beta\cdot\bm j^{(0)})\bigl[\bm{\mathfrak{r}}^{(0)}-\frac{\gamma}{\gamma+1}\bm\beta(\bm\beta\cdot\bm{\mathfrak{r}}^{(0)})\bigr]dV^{(0)}}. \label{meqFLLL} \end{equation} An
averaged time derivative of a quantity varying within a finite
interval is equal to zero. Therefore,
$$\left\langle\frac{d}{dt}\left(\rho^{(0)}\mathfrak{r}_i^{(0)}x_j^{(0)}dV^{(0)}\right) \right\rangle=\frac1c\left\langle\left(\mathfrak{r}_i^{(0)}j_j^{(0)}+\mathfrak{r}_j^{(0)}j_i^{(0)}\right)dV^{(0)} \right\rangle=0  $$
and
$$\left\langle \mathfrak{r}_i^{(0)}j_j^{(0)}dV^{(0)} \right\rangle=\frac12e_{ijk}
\left\langle\left[\bm{\mathfrak{r}}^{(0)}\times\bm j^{(0)}\right]_kdV^{(0)} \right\rangle.  $$
As a result of this transformation, Eq. (\ref{meqFLLL}) takes the form \cite{MLCLE}
\begin{equation}
\bm d=\frac{1}{2c}\int{\left[\bm\beta\times[\bm{\mathfrak{r}}^{(0)}\times\bm j^{(0)}]\right]dV^{(0)}}
=\bm \beta\times\bm \mu^{(0)}. \label{meqFinL} \end{equation}
The motion of a magnetic dipole leads to an appearance of an electric dipole in the lab frame but the total electric charge is equal to zero in any frame.

The validity of the two first terms in the right hand side of Eq. (\ref{deqfinal}) can be easily confirmed with Eq. (\ref{eqpapYF}).
Thus, different methods lead to the same formulas defining the transformation of the electric and magnetic dipole moments from the instantaneously accompanying frame to the lab one.

A possibility to construct a four-tensor from the electric and magnetic dipole moments was first mentioned by Frenkel \cite{Frenkelbook}. However, his assumption that this four-tensor has the form ${\cal D}^{\mu\nu}=(\bm d,\bm \mu)$ [cf. Eq. (\ref{eqKLnmn})] had resulted in incorrect transformation laws of the quantities $\bm d$ and $\bm\mu$. Similar error has been made by Nyborg \cite{Nyborg}. Thus, a correct analysis has not be done in Refs. \cite{Frenkelbook,Nyborg}.

To describe spin effects, we need to express the intrinsic dipole moments in the particle rest frame in terms of the spin (pseudo)vector. In this case, $\bm d_0=d\bm\zeta/s,~\bm\mu_0=\mu\bm\zeta/s$, where $s=|\bm\zeta|$. The quantities $\bm\zeta$ and $s$ have the dimensionality of the angular momentum. The quantity $s$ in classical physics corresponds to $\hbar s$ ($s$ is here the spin quantum number) in quantum mechanics. The dynamics of the dipole moments in the particle rest frame and in the instantaneously accompanying frame differs owing to the Thomas effect. Let us determine a part of the Hamiltonian (\ref{multexp}) which is not conditioned by this effect. It has the form
\begin{equation}\begin{array} {c}
H=-\frac{\mu}{s}\left[\bm B\cdot\bm\zeta-\frac{\gamma}{\gamma+1}(\bm\beta\cdot\bm B)(\bm\beta\cdot\bm\zeta)-(\bm\beta\times\bm E)\cdot\bm \zeta\right]\\-\frac{d}{s}\left[\bm E\cdot\bm\zeta-\frac{\gamma}{\gamma+1}(\bm\beta\cdot\bm E)(\bm\beta\cdot\bm\zeta)+(\bm\beta\times\bm B)\cdot\bm\zeta\right].
\end{array}\label{emgapHn}\end{equation}

It is important that this expression for the Hamiltonian can be reduced with the use of the fields in the instantaneously accompanying frame, $\bm E^{(0)}$ and $\bm B^{(0)}$, satisfying the relations (see Ref. \cite{Jackson})
\begin{equation}\begin{array} {c}
\bm E^{(0)}=\gamma\left[\bm E-\frac{\gamma}{\gamma+1}\bm\beta(\bm\beta\cdot\bm E)
+\bm \beta\times\bm B\right],\\
\bm B^{(0)}=\gamma\left[\bm B-\frac{\gamma}{\gamma+1}\bm\beta(\bm\beta\cdot\bm B)
-\bm \beta\times\bm E\right].
\end{array} \label{meffinal} \end{equation}
Therefore, the Hamiltonian can be presented as follows:
\begin{equation}   \begin{array}{c}
H=-\frac{d\bm E^{(0)}\cdot\bm\zeta}{s\gamma}-\frac{\mu\bm B^{(0)}\cdot\bm\zeta}{s\gamma}.
\end{array} \label{multred} \end{equation}
The use of the Poisson brackets allows one to obtain the corresponding angular velocity of spin precession:
\begin{equation}   \begin{array}{c}
\bm\omega=-\frac{d\bm E^{(0)}}{s\gamma}-\frac{\mu\bm B^{(0)}}{s\gamma}.
\end{array} \label{multanv} \end{equation}

Equations (\ref{emgapHn}) and (\ref{multred}) for the Hamiltonian and Eq. (\ref{multanv}) for the angular velocity of spin precession \emph{in the lab frame} are general.
The angular velocity of spin precession \emph{in the instantaneously accompanying frame} can be obtained from Eq. (\ref{multanv}) with $\gamma\rightarrow1$.

However, we have started from the instantaneously accompanying frame while the spin (pseudo)vector $\bm\zeta$ is defined in the \emph{noninertial} particle rest frame. Angular velocities of spin precession in the two frames differ due to the famous Thomas effect which should also be taken into consideration.

\section{General derivation with allowance for the Thomas effect} \label{allThoeff}

The Thomas effect \cite{Thomas} consists in a change of the angular velocity of spin precession due to a rotation of the particle rest frame. Thomas has shown \cite{Thomas} that the difference between the spin precession in the nonrotating instantaneously accompanying frame and in the rest frame is defined by
\begin{equation}   \begin{array}{c}
\left(\frac{\partial\bm\zeta}{\partial t}\right)_{nonrot}=\left(\frac{\partial\bm\zeta}{\partial t}\right)_{rest\,frame}+\bm\omega_T\times\bm\zeta,
\end{array} \label{Thomprecession} \end{equation}
where $\omega_T$ is the angular velocity of the Thomas precession:
\begin{equation}\begin{array} {c}
\bm\omega_T=-\frac{\gamma^2}{\gamma+1}\left(\bm \beta\times\frac{d\bm \beta}{dt}\right).
\end{array}\label{Thompre}\end{equation}
A very short and clear derivation of Eq. (\ref{Thompre}) has been recently given in Ref. \cite{DraganThomas}.

We can immediately find the total angular velocity of spin precession of the particle with the AMM and EDM with Eqs. (\ref{Thomprecession}) and (\ref{Thompre}).
However, it is more consistent to keep the fully covariant approach.

An inclusion of the EDM brings the covariant equation of spin motion to the general form \cite{FukuyamaSilenko}
\begin{equation}
\frac{d a^\mu}{d\tau}=A_1 F^{\mu\nu}a_\nu+A_2\beta u^\mu F^{\nu\lambda}u_\nu a_\lambda+A_3 G^{\mu\nu}a_\nu+A_4 u^\mu G^{\nu\lambda}u_\nu a_\lambda,
\label{BMT} \end{equation}
where the four-vectors of spin and velocity are defined by
\begin{equation}
a^\mu=(a^0,~{\bm a}),~~~{\bm a}=\bm{\zeta}+\frac{\gamma^2{\bm\beta}({\bm\beta}\cdot\bm{\zeta})}{\gamma+1},~~~a^0={\bm\beta}\cdot{\bm a}=\gamma{\bm\beta}\cdot\bm{\zeta},~~~u^\mu=(\gamma,\gamma{\bm\beta})
\label{spin} \end{equation}
and $G^{\mu\nu}=\epsilon^{\mu\nu\alpha\beta}F_{\alpha\beta}/2$ is the antisymmetric four-tensor dual to the electromagnetic field tensor $F_{\alpha\beta}$. The translational motion of the particle is given by
\begin{equation}
\frac{d u^\mu}{d\tau}=\frac{e}{mc} F^{\mu\nu}u_\nu.
\label{particl} \end{equation}

The transition to the instantaneously accompanying frame and the use of the orthogonality condition $a_\mu u^\mu=0$ result in (see Ref. \cite{FukuyamaSilenko})
\begin{equation}
 A_1=\frac{\mu}{s}, ~~~ A_2=-\frac{1}{s}\left(\mu-\frac{es}{mc}\right)=-\frac{\mu'}{s}, ~~~ A_3=-\frac{d}{s}, ~~~ A_4=\frac{d}{s}.
\label{oeffi} \end{equation}
With the use of Eq. (\ref{particl}), the obtained equation can be presented in the form
\begin{equation}
\frac{da^\mu}{d\tau}=\frac{\mu}{s}\left(F^{\mu\nu}a_\nu-u^\mu F^{\nu\lambda}u_{\nu}a_\lambda\right)-\frac{d}{s}\left(G^{\mu\nu}a_\nu-u^\mu G^{\nu\lambda}u_\nu a_\lambda\right)-
u^\mu \frac{du^\lambda}{d\tau}a_\lambda.
\label{BMT1} \end{equation}

Next derivations can be made similarly to Ref. \cite{Jackson}. It is convenient to denote
\begin{equation}
\Phi^\mu=\frac{\mu}{s}\left(F^{\mu\nu}a_\nu-u^\mu F^{\nu\lambda}u_{\nu}a_\lambda\right)-\frac{d}{s}\left(G^{\mu\nu}a_\nu-u^\mu G^{\nu\lambda}u_\nu a_\lambda\right).
\label{Jackson1} \end{equation}
Evidently, $\Phi^\mu=(\Phi^0,\bm \Phi)$ is a four-vector. Since $u_\mu\Phi^\mu=\gamma(\Phi^0-\bm\beta\cdot\bm\Phi)=0$, it satisfies the relation $\Phi^0=\bm\beta\cdot\bm\Phi$. The last term in Eq. (\ref{BMT1}) can be transformed as follows \cite{Jackson}:
\begin{equation}
u^\mu \frac{du^\lambda}{d\tau}a_\lambda=-u^\mu \gamma\bm a\cdot\frac{d\bm\beta}{d\tau}.
\label{Jackson2} \end{equation}
Thus, Eq. (\ref{BMT1}) leads to
\begin{equation}
\frac{da^0}{d\tau}=\Phi^0+\gamma^2\bm a\cdot\frac{d\bm\beta}{d\tau}, ~~~ \frac{d\bm a}{d\tau}=\bm\Phi+\gamma^2\bm\beta\left(\bm a\cdot\frac{d\bm\beta}{d\tau}\right).
\label{Jackson3} \end{equation}

Now we can calculate the equation of motion for the rest frame spin $\bm\zeta$ with the use of the relations
$$ 
{\bm \zeta}=\bm{a}-\frac{\gamma}{\gamma+1}{\bm\beta}({\bm\beta}\cdot\bm{a}),~~~\frac{d}{d\tau}\left(\frac{\gamma}{\gamma+1}{\bm\beta}\right)=\frac{\gamma}{\gamma+1}\frac{d\bm\beta}{d\tau}+
\frac{\gamma^3}{(\gamma+1)^2}{\bm\beta}\left({\bm\beta}\cdot\frac{d\bm\beta}{d\tau}\right).
$$ 
The needed equation has the form (cf. Ref. \cite{Jackson})
\begin{equation}
\frac{d\bm\zeta}{d\tau}=\bm\Phi-\frac{\gamma\bm\beta}{\gamma+1}\Phi^0+
\frac{\gamma^2}{\gamma+1}\bm\zeta\times\left({\bm\beta}\times\frac{d\bm\beta}{d\tau}\right).
\label{Jackson4} \end{equation}

The transformation of the given four-vector $\Phi^\mu$ to the instantaneously accompanying frame results in
$\bigl(\Phi^{(0)}\bigr)^\mu=\bigl(0,\bm\Phi^{(0)}\bigr)$, where
$$\bm\Phi^{(0)}=\bm\Phi-\frac{\gamma}{\gamma+1}\bm\beta(\bm\beta\cdot\bm\Phi)=\bm\Phi-\frac{\gamma\bm\beta}{\gamma+1}\Phi^0.$$

Since $dt=\gamma \,d\tau$, the derivation of $\bm\Phi^{(0)}$ from Eq. (\ref{Jackson1}) brings the equation of spin motion to the form
\begin{equation}
\frac{d\bm\zeta}{dt}=-\left(\frac{d\bm E^{(0)}}{s\gamma}+\frac{\mu\bm B^{(0)}}{s\gamma}\right)\times\bm\zeta-
\frac{\gamma^2}{\gamma+1}\left({\bm\beta}\times\frac{d\bm\beta}{dt}\right)\times\bm\zeta.
\label{Jackson5} \end{equation}
The angular velocity of spin precession is given by
\begin{equation}
\bm\Omega=-\left(\frac{d\bm E^{(0)}}{s\gamma}+\frac{\mu\bm B^{(0)}}{s\gamma}\right)-
\frac{\gamma^2}{\gamma+1}\left({\bm\beta}\times\frac{d\bm\beta}{dt}\right)=\bm\omega+\bm\omega_T,
\label{angvelo} \end{equation}
where $\bm\omega$ and $\bm\omega_T$ are given by Eqs. (\ref{multanv}) and (\ref{Thompre}), respectively.

Equations (\ref{Jackson5}) and (\ref{angvelo}) show that the total angular velocity of spin precession is the sum of two parts. The first part is given by the Lorentz transformation between the instantaneously accompanying frame and the lab frame. The second one is the contribution from the Thomas precession. This part defines the additional spin precession caused by a purely kinematical effect of a rotation of the particle rest frame (see, e.g., Ref. \cite{Jackson,Rindler}). The presented derivation of Eqs. (\ref{Jackson5}) and (\ref{angvelo}) is fully covariant but it does not specify the two contributions to the total effect. The origins of these contributions are considered in detail in Sec. \ref{Lorentz} and in the theory of the Thomas effect \cite{Thomas,DraganThomas,Jackson,Rindler,Stepanov}.

The particle acceleration is expressed in terms of the lab frame fields as follows:
\begin{equation}
\frac{d\bm\beta}{dt}=\frac{e}{mc\gamma}\left[\bm E+{\bm\beta}\times {\bm B}-\bm\beta(\bm\beta\cdot{\bm E})\right].
\label{eqm} \end{equation}
With the use of Eqs. (\ref{meffinal}), (\ref{eqm}), one can bring Eq. (\ref{angvelo}) to the form
\begin{equation} \begin{array} {c}
\bm \Omega=-\frac{e}{mc}\left[\left(G+\frac{1}{\gamma}\right){\bm B}-\frac{\gamma G}{\gamma+1}({\bm\beta}\cdot{\bm B}){\bm\beta}-\left(G+\frac{1}{\gamma+1}\right)\bm\beta\times{\bm E}\right.\\
+\left.\frac{\eta}{2}\left({\bm E}-\frac{\gamma}{\gamma+1}(\bm\beta\cdot{\bm E})\bm\beta+\bm\beta\times {\bm B}\right)\right],
\end{array} \label{Nelsonh} \end{equation}
where $G=(g-2)/2,~g=2mc\mu/(es)$, and $\eta=2mcd/(es)$. This equation has been previously derived in Ref. \cite{FukuyamaSilenko}.

The equation of spin motion takes a pretty simple form after an expression of the Thomas precession in terms of the fields in the instantaneously accompanying frame. While $$\bm E+{\bm\beta}\times {\bm B}-\bm\beta(\bm\beta\cdot{\bm E})\neq \bm E^{(0)},$$ the angular velocity of the Thomas precession is equal to
\begin{equation}\begin{array} {c}
\bm\omega_T=-\frac{e}{mc(\gamma+1)}\left(\bm \beta\times\bm E^{(0)}\right).
\end{array}\label{Thomiaf}\end{equation}

Therefore,
\begin{equation}   \begin{array}{c}
\frac{d\bm\zeta}{dt}=-\frac{e}{mc}\left(\frac{g\bm B^{(0)}}{2\gamma}+\frac{\eta\bm E^{(0)}}{2\gamma}+\frac{\bm\beta\times\bm E^{(0)}}{\gamma+1}\right)\times\bm\zeta.
\end{array} \label{mufinal} \end{equation}

This final equation explicitly shows the contributions from the electric and magnetic dipole moments and from the Thomas precession.

\section{Discussion and summary}

The earlier derivations of the equation of spin motion of a particle with the AMM and EDM \cite{Nelson,Khriplovich} used the dual transformation of terms proportional to the AMM in the T-BMT equation describing a particle without the EDM. The rigorous derivation of the equation for a particle with the AMM and EDM has been first performed in Ref. \cite{FukuyamaSilenko}. In the present work, we have made a next step and have deduced this equation with the explicit separation of contributions from the Lorentz transformation between the instantaneously accompanying frame and the lab frame and from the Thomas effect. This deduction is fully covariant. The transition to the fields in the instantaneously accompanying frame has allowed us to present the final equation in the very simple form. Amazingly, one need not to divide the magnetic moment into the normal and anomalous parts. This division is a result of expression of the equation of spin motion in terms of the lab frame fields.

In fact, the obtained equation of motion (\ref{mufinal}) cannot exhaustively specify origins of the two contributions. However, a needed specification of the Thomas term is presented by the theory of the Thomas effect \cite{Thomas,DraganThomas,Jackson,Rindler,Stepanov}. This theory shows that the Thomas effect has a purely kinematical origin and is caused by a rotation of the particle rest frame \cite{Jackson,Rindler}. The origin of the contributions of the electric and magnetic dipole moments to Eq. (\ref{mufinal}) has been cleared in Sec. \ref{Lorentz}. It has been demonstrated that the form of these contributions is conditioned by the Lorentz transformations of the electric and magnetic dipole moments from the instantaneously accompanying frame to the lab frame.

These transformations are the key moments of the analysis fulfilled in Sec. \ref{Lorentz}. We have determined the connection between the dipole moments in the lab frame and in the instantaneously accompanying one and
have corrected errors made in previous investigations \cite{Frenkelbook,Nyborg}. We have derived this connection from basic equations of classical electrodynamics, namely, from the Maxwell equations in matter and from the Lorentz transformations of the four-current $j^\mu$ and other four-vectors.
We have also constructed a four-tensor from the electric and magnetic dipole moments.

The above-mentioned new results have been obtained thank to the inclusion of the EDM into the consideration. The existence of the direct relations between the particle spin motion and basic equations of classical electrodynamics is an important fact.
The results obtained show a deep self-consistency of classical electrodynamics.

\section*{Acknowledgements}

The author is grateful to the referees of the journal for valuable remarks and
propositions.
The author also acknowledges the support by the Belarusian Republican Foundation for Fundamental Research
(Grant No. $\Phi$14D-007) and
by the Heisenberg-Landau program of the German Ministry for
Science and Technology (Bundesministerium f\"{u}r Bildung und
Forschung).


\end{document}